\documentclass[superscriptaddress,aps,pra,twocolumn]{revtex4}
\usepackage{bm}
\usepackage{ulem}
\usepackage{epsfig}
\usepackage{graphicx}
\usepackage{amssymb,amsmath,amsbsy,amsgen,amsfonts}    
\usepackage{dcolumn}
\usepackage{amsthm}
\usepackage{mathrsfs}
\usepackage{latexsym}
\usepackage{array}
\usepackage{color}
\usepackage{amstext}
\allowdisplaybreaks[1]
\usepackage{txfonts}
\usepackage{pifont}

\usepackage{epstopdf} 

\newcommand{\abs}[1]{\left\vert #1\right\vert}

\newcommand{\bra}[1]{\left\langle{#1}\right\vert}
\newcommand{\ket}[1]{\left\vert{#1}\right\rangle}

\newcommand{\be}{\begin{equation}}
\newcommand{\ee}{\end{equation}}
\newcommand{\ba}{\begin{array}}
\newcommand{\ea}{\end{array}}
\newcommand{\bqa}{\begin{eqnarray}}
\newcommand{\eqa}{\end{eqnarray}}

\DeclareSymbolFont{symbols}{OMS}{cmsy}{m}{n}

\begin{document}
\title{Quantum noise reduction in intensity-sensitive surface plasmon resonance sensors}

\author{Joong-Sung Lee}
\affiliation{Department of Physics, Hanyang University, Seoul, 04763, Korea}

\author{Trung Huynh}
\affiliation{Department of Physics, Hanyang University, Seoul, 04763, Korea}

\author{Su-Yong Lee}
\affiliation{School of Computational Sciences, Korea Institute for Advanced Study, Seoul, 02455, Korea}

\author{Kwang-Geol Lee}
\affiliation{Department of Physics, Hanyang University, Seoul, 04763, Korea}

\author{\\Jinhyoung~Lee}
\affiliation{Department of Physics, Hanyang University, Seoul, 04763, Korea}

\author{Mark Tame}
\address{School of Chemistry and Physics, University of KwaZulu-Natal, Durban 4001, South Africa}
\address{National Institute for Theoretical Physics, University of KwaZulu-Natal, Durban 4001, South Africa}

\author{Carsten Rockstuhl}
\affiliation{Institute of Theoretical Solid State Physics, Karlsruhe Institute of Technology, 76131 Karlsruhe, Germany}
\affiliation{Institute of Nanotechnology, Karlsruhe Institute of Technology, 76021 Karlsruhe, Germany}

\author{Changhyoup Lee}
\email{changdolli@gmail.com}
\affiliation{Institute of Theoretical Solid State Physics, Karlsruhe Institute of Technology, 76131 Karlsruhe, Germany}

\date{\today}

\begin{abstract}
We investigate the use of twin-mode quantum states of light with symmetric statistical features in their photon number for improving intensity-sensitive surface plasmon resonance (SPR) sensors. For this purpose, one of the modes is sent into a prism setup where the Kretschmann configuration is employed as a sensing platform and the analyte to be measured influences the SPR excitation conditions. This influence modifies the output state of light that is subsequently analyzed by an intensity-difference measurement scheme. We show that quantum noise reduction is achieved not only as a result of the sub-Poissonian statistical nature of a single mode, but also as a result of the non-classical correlation of the photon number between the two modes. When combined with the high sensitivity of the SPR sensor, we show that the use of twin-mode quantum states of light notably enhances the estimation precision of the refractive index of an analyte. With this we are able to identify a clear strategy to further boost the performance of SPR sensors, which are already a mature technology in biochemical and medical sensing applications. 
\end{abstract}


\maketitle
\section{introduction}
A surface plasmon resonance (SPR) is a charge density oscillation resonantly coupled to an electromagnetic field at the interface between a metal and a dielectric. Due to its high field confinement it is extremely sensitive to minute changes in its optical environment~\cite{Raether88}. This feature has prompted its consideration as a sensor in sensing applications, with the potential to achieve a much higher sensitivity when compared to conventional photonic sensors. Furthermore, due to the strong localization of the electromagnetic field to lengths below the diffraction limit, sensing in nanometric spatial domains near the metallic interface becomes possible~\cite{Homola99a, Lal07, Anker08}. Various SPR sensing platforms have been proposed so far, and even commercial products exist~\cite{Rothenhausler88, Jorgenson93, Homola99b, Dostalek05, Sepulveda06, Leung07, Svedendahl09, Mayer11}. Despite practical uses in medical or biochemical science, the ultimate sensitivity of conventional SPR sensors is shot-noise limited due to the intrinsic corpuscular nature of light~\cite{Ran06, Piliarik09, Wang11}. The sensitivity can be improved in general by increasing the input power, but it always remains shot-noise limited. Moreover, the indefinite increase in optical power naturally has its limitations and indeed is very often prohibitive as an excessive increase may cause optical damage to specimens under investigation~\cite{Neuman99, Peterman03, Taylor15, Taylor16} or other unwanted thermal effects~\cite{Kaya13}. Therefore, the allowed maximum input power sets the ultimate sensitivity for plasmonic sensors when employing a classical coherent source. 

Such limited sensitivity imposed by the shot-noise can be surpassed by the use of quantum states of light that have non-Poissonian photon-number distribution or non-classical correlations~\cite{Giovannetti04}. The basic idea for this was first introduced by Caves~\cite{Caves80, Caves81}, a pioneer of a new scientific field called quantum metrology~\cite{Boto00, Giovannetti06, Giovannetti11}, consequently attracting enormous interest from many scientific areas for different purposes~\cite{Maze08, Giovannetti11, Napolitano11, McGuinness11, Steinlechner13, Wolfgramm13, Taylor13}. At present, studies have moved beyond the basic working principles of quantum metrology towards more realistic application scenarios, requiring a robustness of the sensor against losses and any other possible imperfections~\cite{Dorner09, Kacprowicz10, Escher11, Chaves13, Kessler14, Lu15, Zhang15, Jachura16, Ulanov16, Matthews16, Oszmaniec16}. The idea of using quantum resources for sensing or imaging recently inspired efforts to show that both quantum and plasmonic resonance features can be combined for further enhancing the sensing performance of optical devices~\cite{Kalashnikov14, Fan15, Pooser16, Lee16}. It was theoretically demonstrated that quantum plasmonic sensors can overcome both the shot-noise and diffraction limits in plasmonic nanowire interferometric setups~\cite{Lee16}. However, despite the considerable potential of merging plasmonic sensors with quantum techniques~\cite{Tame13}, much work is still to be done in terms of its practical application and understanding the performance of sensors using different types of resource states. Here, various issues arise, such as input and output coupling efficiencies and the need for high interferometric stability. Moreover, various types of classical plasmonic sensors need to be re-examined with quantum resources and measurements.
 
\begin{figure}[t]
\centering
\includegraphics[width=8.7cm]{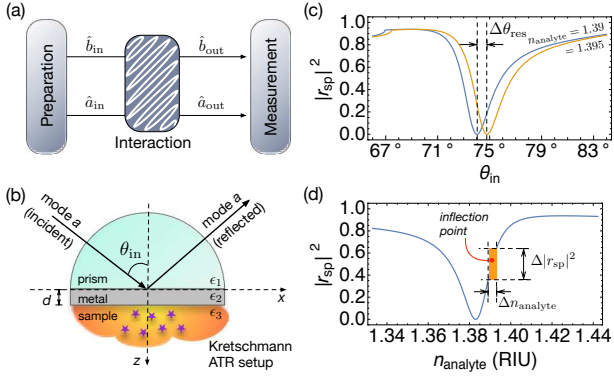}
\caption{
(a) A two-mode sensing scheme that consists of three parts: a preparation, an interaction, and a measurement part.
(b) The Kretschmann configuration for the SPR sensor to be integrated in the interaction part of mode $a$.  
(c) The intensity of reflected light in mode $a$ as a function of incident angle $\theta_{\rm in}$ for two different example analytes: $n_{\rm analyte}=1.39$ (blue curve) and $n_{\rm analyte}=1.395$ (orange curve). A change of the analyte can be identified by observing a shift of the resonance angle, $\Delta \theta_{\rm res}$. Details of the used parameters are provided in the main text. 
(d) The intensity of the reflected light in mode $a$ as a function of the refractive index $n_{\rm analyte}$ (RIU) for a fixed angle of incidence, $\theta_{\rm in}=73^{\circ}$. An orange square box centered at an inflection point determines the most sensitive sensing region for a given incident angle.
}
\label{setup} 
\end{figure}

In this work we investigate in detail the performance of an intensity-sensitive SPR sensor with the use of twin-mode beams that have symmetric statistical features in their photon number. For the SPR sensor, we employ the attenuated-total-reflection (ATR) prism setup using the Kretschmann configuration, the most widely used plasmonic sensing platform that has shown to be implementable with current technology and led to the huge success in commercialization of classical biosensing. The outgoing beams are analyzed using an intensity-difference measurement between the two modes. The general scheme we study is illustrated conceptually in Fig.~\ref{setup}(a). Using this scheme we show that quantum noise reduction can be achieved by exploiting the non-classical statistical features of the prepared twin-mode beams. Eventually, this reduction enhances the estimation precision of the refractive index of an analyte deposited on the metal surface of the ATR setup supporting surface plasmon excitations, as shown in Fig.~\ref{setup}(b). In this scenario we explicitly show how quantum plasmonic sensing exploits both the high sensitivity provided by the SPR and the noise reduction given by the use of quantum resource states. This work constitutes a different quantum plasmonic sensing scheme compared to what was proposed in Ref.~\cite{Lee16}, where the maximum photon number of the input state was restricted to be finite, and the nanowire setup is phase-sensitive, as opposed to the intensity-sensitive scenario considered here.

The paper is structured as follows. In Sec.~II, we provide the quantum mechanical description of the two-mode sensing scheme under investigation. We also characterize the intensity-sensitive sensing behavior using a conventional figure-of-merit, for which input twin-mode beams and an intensity-difference measurement scheme are employed. In Sec.~III, we then use our theoretical framework to show quantum enhancement with exemplarily chosen quantum states. We also provide an understanding of the quantum enhancement obtained using general twin-mode beams. Finally, in Sec. IV, we summarize our work and conclude with an outlook on future studies.

\section{Theoretical description}
We consider a two-mode sensing scheme for measuring the refractive index $n_{\rm analyte}$ of an analyte. As depicted in Fig.~\ref{setup}(a), our two-mode sensor consists of three parts: a preparation part where a two-mode (quantum or classical) state of light is generated, an interaction part where the two-mode light interacts with the analyte, and a measurement part in which the output state of the light is analyzed. Through the interaction part, the input state of light is transformed to the output state. This is described in the Heisenberg picture in terms of the relation between input and output operators, written as
\begin{align}
\hat{a}_{\rm out}&=t_{a}\hat{a}_{\rm in}+r_{b}\hat{b}_{\rm in}+\hat{f}_{a},\label{inout1} \\
\hat{b}_{\rm out}&=r_{a}\hat{a}_{\rm in}+t_{b}\hat{b}_{\rm in}+\hat{f}_{b},\label{inout2}
\end{align}
where the overall transmission ($t$) and reflection ($r$) coefficients can be found via classical electromagnetic theory. The quantum nature of light is included via the operators of the respective electromagnetic field modes. The noise operators $\hat{f}_{a,b}$ take into account losses that occur in the interaction part (e.g., transmission loss through an optical path or absorption when interacting with an analyte), and consequently enable the output operators to preserve the necessary commutation relations~\cite{Barnett96, Barnett98}. 

In the above two-mode scheme, the SPR sensor is implemented in mode $a$ in the interaction part, for which we employ the ATR prism setup using the Kretschmann configuration shown in Fig.~\ref{setup}(b). The highly sensitive response of the surface plasmon excitation on the metal surface for an incident TM-polarized mode is manifested in the reflection coefficient $r_{\rm sp}$ given by~\cite{Raether88}
\begin{align}
r_{\rm sp}
=\abs{r_{\rm sp}}e^{i\phi_{\rm sp}}
=\frac{e^{ i 2 k_{2z} d} r_{23}+ r_{12}}{e^{ i 2 k_{2z} d} r_{23}r_{12} + 1}.\label{rsp}
\end{align}
Here, $r_{lm}=\left[\frac{k_{lz}}{\epsilon_{l}} - \frac{k_{mz}}{\epsilon_{m}}\right]\Big/\left[\frac{k_{lz}}{\epsilon_{l}}+\frac{k_{mz}}{\epsilon_{m}}\right]$ for $l,m\in \{1,2,3 \}$, $k_{lz}=\cos\theta_{\rm in}\sqrt{\epsilon_{l}}\omega/c$ denotes the normal-to-surface component of the wave vector in the $l$th medium, $\epsilon_{l}$ is the respective permittivity, and $d$ is the thickness of the second layer. As shown in Fig.~\ref{setup}(b), the Kretschmann configuration consists of three layers: $\epsilon_{1}=n_{\rm prism}^{2}$, $\epsilon_{2}=\epsilon_{\rm metal}(\omega)$, and $\epsilon_{3}=n_{\rm analyte}^{2}$, where the analyte to be measured is deposited on the metal film. The highly sensitive behavior of the reflectance $\vert r_{\rm sp}\vert^{2}$ to the variation in $n_{\rm analyte}$ is shown in Figs.~\ref{setup}(c) and (d), where we choose $n_{\rm prism}=1.5107$, a wavelength $\lambda_{0}=810~{\rm nm}$, and a gold film with $d=50~{\rm nm}$, whose dielectric function $\epsilon_{\rm metal}(\omega)$ is given by experimental data~\cite{Rakic98}. These ATR parameters will be used throughout the analysis of this work. When the refractive index unit (RIU) of the analyte changes, a shift of the resonance angle $\theta_{\rm res}$ at which the ATR occurs is observed. This can be identified by analyzing the intensity of the reflected mode $a$ as a function of the incident angle $\theta_{\rm in}$ [see Fig.~\ref{setup}(c)], or alternatively the change of the reflected intensity is observed for a fixed angle of incidence [see Fig.~\ref{setup}(d), where we use $\theta_{\rm in}=73^{\circ}$ for example]. For the latter, we consider $n_{\rm analyte}$ in a range from $1.333$ to $1.4422$, corresponding to a water solution that contains Bovine Serum Albumin with the concentration varying from $0\%$ to $60\%$~\cite{Leung07, Lee16}. A square box in Fig.~\ref{setup}(d) represents the most sensitive sensing region defined by the inflection point $n_{\rm analyte}^{(\rm inf)}$. There, an infinitesimal change of the refractive index can be recognized by observing the largest possible change of reflected intensities. We will particularly focus on this region throughout this work as an example of typical intensity-sensitive sensing schemes~\cite{Homola99a}. 

We treat the ATR prism setup using an effective quantum beam splitter model~\cite{Tame08, Ballester09}, so that the transmission coefficient $t_{a}$ includes the reflection coefficient of Eq.~(\ref{rsp}) that contains the information about concentration changes in the sample we aim to measure. For a realistic situation, channel losses are also taken into account in our theoretical description using the fictitious beam splitter model with the transmission coefficients $\eta_{a}$ and $\eta_{b} \in \mathbb{R}$ for mode $a$ and $b$, respectively~\cite{Loudonbook}. Metallic Ohmic losses associated with the excitation of the SPR are considered by the imaginary part of the dielectric function of the metal $\epsilon_{\rm metal}(\omega)$, causing a broadening in $\vert r_{\rm sp}\vert$. Additional optical components (e.g., a beam splitter) can be inserted in the interaction part, but here we focus on the simplest case that the overall transmission and reflection coefficients in Eqs.~(\ref{inout1}) and (\ref{inout2}) are written as $t_{a}=e^{i(\phi_{\rm sp}+\theta)} \abs{r_{\rm sp}}\eta_{a}$, $t_{b}=\eta_{b}$, and $r_{a}= r_{b}=0$, where $\theta$ is the relative phase caused by different travel lengths $\Delta L$ between mode $a$ and $b$. That is, mode $a$ interacts with the analyte while mode $b$ is kept as a reference. 

From this input-output relation it is clear that the transmitted light with coefficient $t_{a}$ (or both $t_{a}$ and $t_{b}$ in general) exhibits the behaviors shown in Fig.~\ref{setup}(d). However, the actual output signal and its associated noise are obtained by applying the input-output relations of Eqs.~(\ref{inout1}) and (\ref{inout2}) to a given input state. The initial statistical feature of the input state of light thus affects the output signal. It constitutes the fundamental cause that blurs the signal curves in Figs.~\ref{setup}(c) and (d), eventually making it hard to perceive the changes, e.g., $\Delta \theta_{\rm res}$ or $\Delta \vert r_{\rm sp}\vert^{2}$, when an infinitesimal variation in $n_{\rm analyte}$ occurs. 

We concentrate on the use of input twin-mode beams that have equal average photon numbers as well as equal fluctuations, {\it i.e.}, $\langle \hat{n}_{a_{\rm in}}\rangle = \langle \hat{n}_{b_{\rm in}}\rangle=N$ and $\langle \Delta \hat{n}_{a_{\rm in}}\rangle = \langle \Delta \hat{n}_{b_{\rm in}}\rangle$ where $\hat{n}_{a_{\rm in} (b_{\rm in})}=\hat{a}^{\dagger}_{\rm in}\hat{a}_{\rm in} (\hat{b}^{\dagger}_{\rm in}\hat{b}_{\rm in})$, $\langle \Delta \hat{O}\rangle=[ \langle \hat{O}^{2}\rangle -\langle \hat{O}\rangle^{2} ]^{1/2}$ denotes the standard deviation of the operator $\hat{O}$, and $\langle .. \rangle$ denotes the expectation value with respect to an input state. Twin-mode states can be written in the Fock state basis $\{\ket{n,m}\}$ as $\ket{\psi_{\rm twin}}=\sum_{n,m=0}^{\infty}C_{n,m}\ket{n,m}_{ab}$ with $\sum_{n,m=0}^{\infty}\abs{C_{n,m}}^{2}=1$ and the twin-mode condition $\abs{C_{n,m}}=\abs{C_{m,n}}$. This state includes the class of path-symmetric states, for which $C_{n,m}=C_{m,n}^{*}e^{-2i\chi_{0}}$ needs to be satisfied with a constant phase factor $\chi_{0}$~\cite{Hofmann09, Seshadreesan13}. Using the twin-mode states we focus on the corpuscular nature of light, e.g., non-classical photon-number distribution or photon-number correlations~\cite{Meda16}. 

After the interaction with an analyte via the SPR sensor in mode $a$, the output state is analyzed by a measurement. For practical relevance, we employ the intensity-difference measurement scheme that was recently used in experiments for quantum plasmonic sensing~\cite{Fan15,Pooser16} and quantum imaging~\cite{Brida10, Brida11}. This measurement, written in terms of the output operators as $\hat{M}=\hat{b}_{\rm out}^{\dagger}\hat{b}_{\rm out}-\hat{a}_{\rm out}^{\dagger}\hat{a}_{\rm out}$, perceives neither the effect of a relative phase $e^{i(\phi_{\rm sp}+\theta)}$ nor the relative phases of the input state, $e^{i{\rm arg}[C_{n,m}]}$, so that our sensor serves as purely intensity-sensitive. Furthermore, the common noise existing in mode $a$ and $b$, or originating from the source can be eliminated by this measurement~\cite{Bachorbook}. The measurement signal and associated noise with respect to the output state can be calculated via Eqs.~(\ref{inout1}) and (\ref{inout2}) for a given input state within the Heisenberg picture. The estimation precision $\delta n_{\rm analyte}$ that characterizes the sensing performance can be obtained by the linear error propagation method~\cite{Durkin07}, 
\begin{equation}
\delta n_{\rm analyte}=\frac{\langle \Delta \hat{M}\rangle}{\left\vert\frac{\partial \langle \hat{M}\rangle}{\partial n_{\rm analyte}}\right\vert},
\label{deltan}
\end{equation}
where $\langle .. \rangle$ denotes here the expectation value with respect to the output state, which is the input state transformed according to Eqs.~(\ref{inout1}) and (\ref{inout2}). For an arbitrary twin-mode input state, the signal is written as $\langle \hat{M} \rangle=(\eta_{b}^{2}-\vert r_{\rm sp}\vert^{2}\eta_{a}^{2}) N$ (see the Appendix for details), whose steepness depending on $n_{\rm analyte}$ at a fixed angle can be maximized when $\eta_{a}=\eta_{b}=\eta$, consequently decreasing $\delta n_{\rm analyte}$. Such balanced losses can be controlled by inserting variable neutral density filters in the optical paths. Moreover, the impact of the detection efficiency can be also accommodated in $\eta$~\cite{Foxbook}. Thus, the measurement signal is directly proportional to the inversion of the reflectance, {\it i.e.}, $\langle \hat{M} \rangle=(1-\vert r_{\rm sp}\vert^{2}) \eta^{2} N$. Note that this depends not on the phase of input state but only on the average photon number of the input state. It is also apparent that the sensitive behavior of $\vert r_{\rm sp}\vert^{2}$ to $n_{\rm analyte}$ (RIU) is directly accommodated in the denominator of Eq.~(\ref{deltan}), {\it i.e.}, $\left\vert \frac{\partial \langle \hat{M}\rangle}{\partial n_{\rm analyte}}\right\vert =\eta^{2}N \left\vert \frac{\partial \vert r_{\rm sp}\vert^{2}}{\partial n_{\rm analyte}}\right\vert$, commonly called sensitivity. For this sensitivity to be maximized for given $N$ and $\eta$, {\it i.e.}, for any small change in $n_{\rm analyte}$ to be perceived by a significant change in $\langle \hat{M}\rangle$, one needs to calibrate the operating region of $n_{\rm analyte}$ close to the inflection point $n_{\rm analyte}^{\rm (inf)}$ where the denominator of Eq.~(\ref{deltan}) is maximized  for a given angle of incidence [see the square box in Fig.~\ref{setup}(d)]. The inflection point $n_{\rm analyte}^{\rm (inf)}$ changes with the incident angle $\theta_{\rm in}$. One can readily tune the sensing region defined as $n_{\rm analyte}^{\rm (inf)}$ through varying the incident angle $\theta_{\rm in}$. We also stress that the classical properties of the SPR are responsible for the sensitivity of a sensor, while any non-classical feature of the input state is responsible for the numerator in Eq.~(\ref{deltan}). In other words, the plasmonic features help increasing the denominator and the quantum features are able to decrease the numerator in our sensing scenario. Thus, such cooperation between the classical SPR and quantum resource enables the estimation precision $\delta n_{\rm analyte}$ to be further improved than that associated with the shot-noise limit. The impact of the quantum resource is elaborated in the next section. 

\section{Quantum enhancement}
To quantify the quantum enhancement, it is useful to define a ratio of the precision to the precision obtained using a classical reference, for which we consider the product coherent state. The ratio ${\cal R}$ is defined using Eq.~(\ref{deltan}) as
\begin{equation}
{\cal R}
=\frac{\delta n_{\rm analyte,(c)}}{\delta n_{\rm analyte}}
=\frac{\langle \Delta \hat{M}\rangle_{\rm (c)}}{\langle \Delta \hat{M}\rangle},
\end{equation}
where the subindex ``(c)'' denotes a calculation with respect to the product coherent state $\ket{\alpha \alpha}_{ab}=e^{-\abs{\alpha}^{2}}\sum_{n,m=0}^{\infty}\frac{\alpha^{n}\alpha^{m}}{\sqrt{n! m!}}\ket{n,m}$ with $\vert \alpha\vert^{2}=N$. The greater-than-unity value (${\cal R}>1$), reveals an enhancement in the estimation precision $\delta n_{\rm analyte}$, or equivalently the quantum noise reduction in $\langle \Delta \hat{M}\rangle$ as compared to the classical reference. For an arbitrary input twin-mode beam, it can be shown that the ratio is characterized by the statistical features of the photon-number distribution and two-mode correlation of the input state, which reads (see the Appendix for details)
\begin{align}
{\cal R}
=\left(
\frac{1+\abs{r_{\rm sp}}^{2}}
{\left(1-\abs{r_{\rm sp}}^{2}\right)^{2} \eta^{2} Q_{\rm M}+2\abs{r_{\rm sp}}^{2}\eta^{2} \sigma +1+\abs{r_{\rm sp}}^{2}(1-2\eta^{2})}
\right)^{1/2},
\label{eq.ratio}
\end{align}
where $Q_{\rm M}$ denotes the Mandel-$Q$ parameter and $\sigma$ denotes the degree of correlation between the two modes, defined as
\begin{align*}
Q_{\rm M}
=\frac{\langle \Delta \hat{n}_{a_{\rm in}}\rangle^{2}}{\langle \hat{n}_{a_{\rm in}}\rangle}-1,
\text{~and~}
\sigma 
=\frac{\langle \Delta(\hat{n}_{b_{\rm in}}-\hat{n}_{a_{\rm in}})\rangle^{2}}{\langle \hat{n}_{a_{\rm in}}\rangle +\langle \hat{n}_{b_{\rm in}}\rangle},
\end{align*}
respectively. It is evident that the ratio ${\cal R}$ is independent of any phase since our sensing scheme is only intensity-sensitive. The enhancement identified by ``${\cal R}>1$'' can be achieved by lowering $Q_{\rm M}$ or $\sigma$ below the values for the product coherent state ({\it i.e.}, $Q_{\rm M}=0$ and $\sigma=1$ for $\ket{\alpha \alpha}_{ab}$). Since a variance is non-negative, the beneficial regimes take the range $-1 \le Q_{\rm M}<0$ and $0\le \sigma<1$, and these regimes are only accessible by non-classical states of light, although not all non-classical states give values in these ranges. Interestingly, $\sigma$ can also be represented in terms of the intermode correlation ${\cal J}={\rm cov}[\hat{n}_{a_{\rm in}},\hat{n}_{b_{\rm in}}]/\langle\Delta\hat{n}_{a_{\rm in}}\rangle\langle\Delta\hat{n}_{b_{\rm in}}\rangle$ \cite{Gerrybook} by $\sigma=(1+Q_{\rm M})(1-{\cal J})$ for an arbitrary input twin-mode beam. Thus, $\sigma$ and $Q_{\rm M}$ are not independent features. Since $\abs{{\cal J}}\le 1$, the states characterized by $Q_{\rm M}<0$ and ${\cal J}>0$ immediately guarantee $\sigma<1$.

From Eq.~(\ref{eq.ratio}), it can be shown that ${\cal R}$ approaches unity, regardless of $\vert r_{\rm sp}\vert$ and the input state, as $\eta$ decreases, {\it i.e.}, the quantum enhancement diminishes as the channel transmission and/or detection efficiency decrease. The behavior of ${\cal R}$ can be seen to be equivalent to a quantum imaging problem with an absorbing object \cite{Brambilla08} if the ATR represented by a reduction in $\abs{r_{\rm sp}}$ is treated as an absorption. This implies that the estimation of $\abs{r_{\rm sp}}$ is a crucial step in intensity-sensitive SPR sensing to estimate the refractive index of analyte with better precision. Despite the mathematical equivalence with quantum imaging, SPR sensing requires a proper investigation with the use of quantum resources since the ATR depends on many different parameters, e.g., $\theta_{\rm in}$, $\lambda_{0}$, $d$, and $n_{\rm prism}$.

\begin{figure}[t]
\centering
\includegraphics[width=8.7cm]{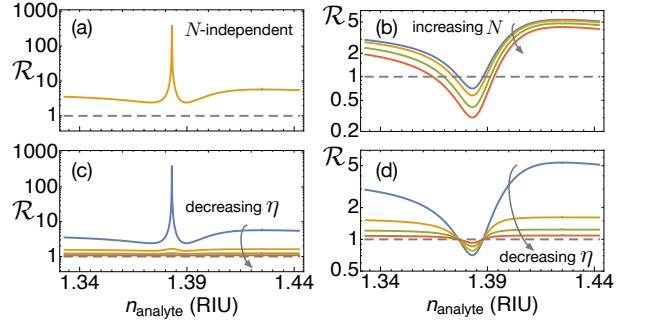}
\caption{Behaviors of ${\cal R}$ for the twin Fock state (left column) and TMSV state (right column) with increasing average photon number ($N=1,2,5$, and $10$) (upper row) and decreasing channel transmission and/or detection efficiency ($\eta=1, 0.8, 0.6, 0.4$) (lower row). The dashed gray line at ${\cal R}=1$ denotes the classical reference, above which a quantum enhancement is observed. 
}
\label{ratio} 
\end{figure}

\subsection{Examples of twin-mode beam states}
Now let us look at particular behaviors of the ratio ${\cal R}$ for typical examples of twin-mode beams. First, it is apparent from Eq.~(\ref{eq.ratio}) that the best twin-mode beam is the twin Fock state $\ket{N,N}_{ab}=\frac{(\hat{a}_{\rm in}^{\dagger})^{N}}{(N!)^{1/2}}\frac{(\hat{b}_{\rm in}^{\dagger})^{N}}{(N!)^{1/2}}\ket{0,0}$ for which the Mandel-$Q$ parameter and the degree of correlation attain minimal values, {\it i.e.}, $Q_{\rm M}=-1$ and $\sigma=0$, so that ${\cal R}_{NN}=\left[(1+\abs{r_{\rm sp}}^{2})/(\abs{r_{\rm sp}}^{2}-\abs{r_{\rm sp}}^{4})\right]^{1/2}$ when $\eta=1$. It is interesting that the ratio ${\cal R}$ is independent of the photon number $N$, which indicates that the twin single-photon state $\ket{11}_{ab}$ is sufficient to achieve the same amount of enhancement compared to the $N$-photon state case. While relying on the same reason, the twin single-photon state has very recently been used in absorption spectroscopy \cite{Whittaker17}. One can also observe that ${\cal R}$ diverges when $\abs{r_{\rm sp}}^{2}\rightarrow 1$ or $0$, since the photon-number fluctuation of the output state for the input twin Fock state vanishes in these two limits, whereas that of the input product coherent state is kept finite.

Another interesting twin-mode beam that exhibits quantum enhancement is the two-mode squeezed vacuum (TMSV) state defined as $\ket{\rm TMSV}=\hat{S}_{ab}(\xi)\ket{0,0}_{ab}$, where $\hat{S}_{ab}(\xi)=\exp[\xi^{*}\hat{a}_{\rm in}\hat{b}_{\rm in}-\xi\hat{a}_{\rm in}^{\dagger}\hat{b}_{\rm in}^{\dagger}]$ denotes the two-mode squeezing operator with a squeezing parameter $\xi\in\mathbb{C}$. The TMSV state, conventionally called just ``twin beams'', reveals the same correlation as the twin Fock state, {\it i.e.}, $\sigma=0$, but $Q_{\rm M}=N$, so that ${\cal R}_{\rm TMSV}=\left((1+\abs{r_{\rm sp}}^{2})/(1-\abs{r_{\rm sp}}^{2}+N(1-\abs{r_{\rm sp}}^{2})^{2}) \right)^{1/2}$ when $\eta=1$. This shows that the increase of input power, equivalent to increasing $N$, diminishes the quantum enhancement. A significant quantum enhancement can be obtained when $N\ll 1/(1-\abs{r_{\rm sp}}^{2})^{2}$, which is readily achievable with moderate optical power when $\abs{r_{\rm sp}}^{2}\approx1$. On the other hand, as $\abs{r_{\rm sp}}^{2}$ decreases, the enhancement is only possible in the extremely low-photon regime ($N\ll 1$), and then eventually becomes no longer possible when $\abs{r_{\rm sp}}^{2}$ is close to zero, at which point ${\cal R}_{\rm TMSV} \approx(1+N)^{-1/2}$ is always below unity.

In Fig.~\ref{ratio} we present the aforementioned behaviors with varying refractive index from $1.333$ (RIU) to $1.4422$ (RIU) at a fixed angle of incidence, for which we choose $\theta_{\rm in}=73^{\circ}$ as an example. As the incident angle changes, the inflection point $n_{\rm analyte}^{\rm (inf)}$ changes, so that the peaks or dips are shifted with a small change of height, but the general feature remains the same. Figure \ref{ratio}(a) shows the independence of $N$ for the input twin Fock state and also a tremendous peak around $n_{\rm analyte}\approx1.383$ where $\abs{r_{\rm sp}}\approx0$. Although the ratio ${\cal R}$ exhibits a huge enhancement at that point, the corresponding region cannot be used for sensing as the sensitivity is very small, {\it i.e.}, $\abs{\frac{\partial \langle \hat{M} \rangle}{\partial n_{\rm analyte} }}\ll1$, which worsens the estimation precision of Eq.~(\ref{deltan}). The TMSV state, on the other hand, shows a strong dependence of $N$ where the increase of input power significantly decreases the enhancement as shown in Fig.~\ref{ratio}(b), so that the use of a TMSV is only useful in the low-photon regime. Regardless of $n_{\rm analyte}$ and the input state, ${\cal R}\rightarrow 1$ as $\eta \rightarrow 0$, expected from Eq.~(\ref{eq.ratio}), as shown in Fig.~\ref{ratio}(c) for the twin Fock state and Fig.~\ref{ratio}(d) for the TMSV state.

\begin{figure}[t]
\centering
\includegraphics[width=8.7cm]{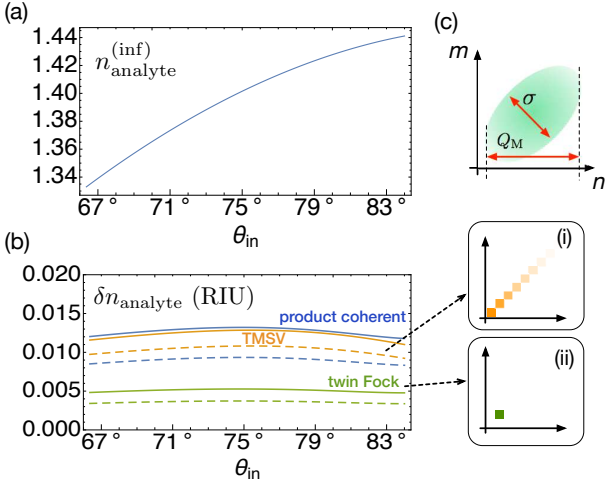}
\caption{For given incidence angles $\theta_{\rm in}$ ranging from $65.5^{\circ}$ to $83.5^{\circ}$, (a) the inflection point of $n_{\rm analyte}^{\rm (inf)}$ is determined, at which the sensitivity is maximized, and (b) the estimation precision at the inflection point can be calculated from Eq.~(\ref{deltan}) for the product coherent, twin Fock and TMSV state, for which we choose $N=1$ (solid lines) and $N=2$ (dashed lines) as examples. (c) Visual quantifications of $Q_{\rm M}$ and $\sigma$ in the distribution of $\abs{C_{n,m}}^{2}$ in the $n$-$m$ plane for an arbitrary twin-mode beam $\ket{\psi_{\rm twin}}$. Insets in (b) also visualize the two-dimensional photon-number distributions for the twin Fock and TMSV states, clearly showing that they have the same broadening along the anti-diagonal direction (yielding the same $\sigma$), but different broadening in the horizontal direction (yielding different $Q_{\rm M}$).
}
\label{inflection}
\end{figure}

The inflection point $n_{\rm analyte}^{\rm (inf)}$ at which the sensitivity is maximized is presented with an incidence angle in a range from $65.5^{\circ}$ to $83.5^{\circ}$ in Fig.~\ref{inflection}(a). Furthermore, the estimation precisions for the product coherent, twin Fock, and TMSV states are calculated at the inflection point given according to $\theta_{\rm in}$. In Fig.~\ref{inflection}(b), we present the results for $N=1$ (solid lines) and $N=2$ (dashed lines). As expected from Fig.~\ref{ratio}, the twin Fock state shows the best estimation precision for both $N=1,2$, while the TMSV state is rather comparable to the product coherent state, {\it i.e.}, better when $N=1$, whereas worse when $N=2$.

\subsection{General twin-mode beam states}
The above examples are particular cases of twin-mode beams. In general, the quantities $Q_{\rm M}\propto \sum_{n,m=0}^{\infty}n^{2}\abs{C_{n,m}}^{2}$ and $\sigma\propto \sum_{n,m=0}^{\infty}(n-m)^{2}\abs{C_{n,m}}^{2}$ for a given $N$ can be understood as the broadenings of a $\abs{C_{n,m}}^{2}$ distribution along the horizontal and anti-diagonal directions in the $n$-$m$ plane, respectively, as shown in Fig.~\ref{inflection}(c). In other words, when the respective broadenings are minimized, the corresponding quantities are minimized. Therefore, the twin Fock state is definitely the best example that shows no broadening in both directions [see inset (i) of Fig.~\ref{inflection}(b)], whereas the TMSV state reveals a broadening along the diagonal axis [see inset (ii) of Fig.~\ref{inflection}(b)]. The latter state exhibits a larger broadening than the Poisson distribution in the horizontal axis and is called a super-Poissonian distribution. From such a point-of-view, keeping in mind ${\cal R}$ is maximized via lowering $Q_{\rm M}$ and $\sigma$, one can immediately conclude that the NOON state that is routinely considered in quantum metrology~\cite{Giovannetti04, Mitchell04, Walther04, Afek10} is worse than the product coherent state, since it has larger broadenings in both directions, {\it i.e.}, $Q_{\rm M}=N-1$ and $\sigma=2N$ for $\ket{\rm NOON}=\frac{1}{\sqrt{2}}\left(\ket{2N,0}+\ket{0,2N}\right)$. However, a NOON state is more favorable when a suitable measurement scheme is employed in the case when there is no loss~\cite{Lee16}. With the same reason, the product squeezed state cannot be a good resource for our intensity-sensitive SPR sensing, {\it i.e.}, $Q_{\rm M}=N+1$ and $\sigma=2N+2$ for $\ket{\xi,\xi}=\hat{S}_{a}(\xi)\otimes\hat{S}_{b}(\xi)\ket{0,0}_{ab}$ where the single mode squeezing operator denotes $\hat{S}_{a}(\xi)=\exp[\frac{1}{2}(\xi^{*}\hat{a}_{\rm in}^{2}-\xi \hat{a}_{\rm in}^{\dagger 2})]$ and the same for mode $b$. Instead, other twin-mode beams, e.g., pair coherent states \cite{Agarwal86} or finite-dimensional photon-number entangled states \cite{Suyong12}, might be more useful, resulting in better sensing performance than the TMSV state since $Q_{\rm M}$ can be reduced below zero while keeping $\sigma=0$. 

The above analysis that takes into account a general twin-mode beam state allows one to predict and compare the sensing performances among different quantum resources only in terms of $Q_{\rm M}$ and $\sigma$. This highlights the use of commonly used typical states and opens up a fruitful direction for future study.  

\section{Remarks}
In this work we studied an intensity-sensitive SPR sensor illuminated with twin-mode beams that have symmetric statistical features in their photon number to enhance the estimation precision. We showed that the non-classical features of photon-number distribution of the input state reduce the quantum noise in the measurement signals and consequently improve the sensing performance of the plasmonic sensor. This constitutes a different quantum plasmonic sensing scheme than has previously been considered by exploiting quantum features not used in the phase-sensitive plasmonic nanowire sensing platform \cite{Lee16}.
We clearly revealed that the high sensitivity is provided by the SPR, while quantum noise reduction is provided by non-classical features of the input state. Thus their cooperation plays an important role in quantum plasmonic sensing. 

As this work focused on an intensity-sensitive SPR sensing scheme, it would also be interesting to study phase-sensitive SPR sensing scenarios~\cite{Wu04, Kashif14}. Furthermore, the sensitivity provided by the SPR in the Kretschmann prism setup can be further enhanced by replacing the metal film by stacks of periodically arranged dielectric layers which support so-called Bloch surface waves that exhibit much sharper resonance curves due to lower loss~\cite{Yeh78, Toma13, Frascella13}. The ultimate fundamental limit to the estimation precision of Eq.~(\ref{deltan}) can be obtained via the Quantum Cram\'er-Rao bound, but we leave this for future study since the calculation of the quantum Fisher information about a parameter characterizing a non-unitary evolution (intensity change) is quite demanding compared to the case of the estimation of a parameter characterizing a unitary evolution (phase change) \cite{Monras07, Escher11, Alipour14}.

\section*{ACKNOWLEDGMENTS}
This work was supported by the South African National Research Foundation and the National Laser Centre, the Basic Science Research Program through the National Research Foundation (NRF) of Korea funded by the Ministry of Science and ICT (No.2016R1A2B4014370 and No.2014R1A2A1A10050117), and the Information Technology Research Center (ITRC) support program (IITP-2016-R0992-16-1017) supervised by the Institute for Information \& communications Technology Promotion (IITP). 

\section*{APPENDIX}
\begin{appendix}
Here we provide the details of calculations for the results given in the main text. First, the expectation value of the intensity difference measurement is calculated as
\begin{align*}
\langle \hat{M} \rangle
&= \bra{\Psi_{\rm out}}\hat{b}_{\rm out}^{\dagger}\hat{b}_{\rm out}\ket{\Psi_{\rm out}}-\bra{\Psi_{\rm out}}\hat{a}_{\rm out}^{\dagger}\hat{a}_{\rm out}\ket{\Psi_{\rm out}}\\
&= \eta_{b}^{2}\bra{\psi_{\rm in}}\hat{b}_{\rm in}^{\dagger}\hat{b}_{\rm in}\ket{\psi_{\rm in}} -\abs{r_{\rm sp}}^{2}\eta_{a}^{2}\bra{\psi_{\rm in}}\hat{a}_{\rm in}^{\dagger}\hat{a}_{\rm in}\ket{\psi_{\rm in}}\\
&= (\eta_{b}^{2} -\abs{r_{\rm sp}}^{2}\eta_{a}^{2})N,
\end{align*}
where $\ket{\Psi_{\rm out}}$ denotes the output state transformed via Eqs.~(\ref{inout1}) and (\ref{inout2}), including the two modes, the associated bath modes, and the surface plasmon mode. Here it is assumed that the initial state of bath modes and surface plasmon mode are the vacuum states. In the same manner, the standard deviation of the intensity difference measurement is calculated as
\begin{widetext}
\begin{align*}
\langle \Delta \hat{M} \rangle
&= \Big[\bra{\Psi_{\rm out}} \Delta \hat{n}_{b_{\rm out}}\ket{\Psi_{\rm out}}^{2} + \bra{\Psi_{\rm out}} \Delta \hat{n}_{a_{\rm out}}\ket{\Psi_{\rm out}}^{2}
-2 \left(\bra{\Psi_{\rm out}} \hat{n}_{b_{\rm out}} \hat{n}_{a_{\rm out}}\ket{\Psi_{\rm out}} 
- \bra{\Psi_{\rm out}} \hat{n}_{b_{\rm out}}\ket{\Psi_{\rm out}} \bra{\Psi_{\rm out}} \hat{n}_{a_{\rm out}} \ket{\Psi_{\rm out}}\right)\Big]^{1/2}\\
&= \Big[\eta_{b}^{4}\bra{\psi_{\rm in}} \Delta \hat{n}_{b_{\rm in}}\ket{\psi_{\rm in}}^{2} +\eta_{b}^{2}\left(1-\eta_{b}^{2}\right)\bra{\psi_{\rm in}} \hat{n}_{b_{\rm in}}\ket{\psi_{\rm in}}
+\abs{r_{\rm sp}}^{4}\eta_{a}^{4}\bra{\psi_{\rm in}} \Delta \hat{n}_{a_{\rm in}}\ket{\psi_{\rm in}}^{2} +\abs{r_{\rm sp}}^{2}\eta_{a}^{2}\left(1-\abs{r_{\rm sp}}^{2}\eta_{a}^{2}\right)\bra{\psi_{\rm in}} \hat{n}_{a_{\rm in}}\ket{\psi_{\rm in}}\\
&~~~~~~~ + \abs{r_{\rm sp}}^{2}\eta_{a}^{2}\eta_{b}^{2}\bra{\psi_{\rm in}}\Delta\left(\hat{n}_{b_{\rm in}}-\hat{n}_{a_{\rm in}}\right)\ket{\psi_{\rm in}}^{2}
-\abs{r_{\rm sp}}^{2}\eta_{a}^{2}\eta_{b}^{2} \bra{\psi_{\rm in}} \Delta\hat{n}_{a_{\rm in}}\ket{\psi_{\rm in}}
-\abs{r_{\rm sp}}^{2}\eta_{a}^{2}\eta_{b}^{2} \bra{\psi_{\rm in}} \Delta\hat{n}_{b_{\rm in}}\ket{\psi_{\rm in}}\Big]^{1/2}\\
&= N^{1/2}
\left[
\left(\eta_{b}^{2}-\abs{r_{\rm sp}}^{2}\eta_{a}^{2}\right)^{2}Q_{\rm M} 
+2\abs{r_{\rm sp}}^{2}\eta_{a}^{2}\eta_{b}^{2} \sigma
+\eta_{b}^{2} +\abs{r_{\rm sp}}^{2}\eta_{a}^{2}\left(1-2\eta_{b}^{2}\right)
\right]^{1/2},
\end{align*}
\end{widetext}
by which the ratio ${\cal R}$ in Eq.~(\ref{eq.ratio}) is straightforwardly obtained. 

\end{appendix}

\end{document}